\documentclass[conference]{IEEEtran}
\IEEEoverridecommandlockouts
\usepackage[whole]{bxcjkjatype}

\usepackage{cite}
\usepackage{amsmath,amssymb,amsfonts}
\usepackage{algorithmic}
\usepackage{graphicx}
\usepackage{textcomp}
\usepackage{xcolor}
\usepackage{caption}

\usepackage[most]{tcolorbox}

\newtcolorbox{promptbox}{
  colback=gray!5,
  colframe=gray!60,
  boxrule=0.5pt,
  arc=2pt,
  left=4pt,
  right=4pt,
  top=4pt,
  bottom=4pt,
  fontupper=\small,
  breakable
}

\usepackage[hyphens]{url}
\usepackage[colorlinks=true, linkcolor=black, urlcolor=black, citecolor=black]{hyperref}
\def\BibTeX{{\rm B\kern-.05em{\sc i\kern-.025em b}\kern-.08em
T\kern-.1667em\lower.7ex\hbox{E}\kern-.125emX}}

\hypersetup{
    colorlinks=true,
    linkcolor=black,
    urlcolor=black,
    citecolor=black,
    breaklinks=true
}

\begin{document}

\title{Supporting Perspective Acquisition and Opinion Formation on Societal Issues through AI-Generated Japanese Rap Battle Debates}

\author{
\IEEEauthorblockN{
Ryota Mibayashi\IEEEauthorrefmark{1},
Toru Urakawa\IEEEauthorrefmark{2},
Dai Takanashi\IEEEauthorrefmark{3},
Tomoya Morohoshi\IEEEauthorrefmark{3},
Kanata Yamagishi\IEEEauthorrefmark{3},\\
Ryuho Sekikawa\IEEEauthorrefmark{4},
Yasuhiko Nishimura\IEEEauthorrefmark{5},
Yuta Takeuchi\IEEEauthorrefmark{3},
Hideaki Tamori\IEEEauthorrefmark{2},\\
Takehiro Yamamoto\IEEEauthorrefmark{6},
Hidenari Kiyomitsu\IEEEauthorrefmark{1},
and Hiroaki Ohshima\IEEEauthorrefmark{6}
}

\IEEEauthorblockA{
\IEEEauthorrefmark{1}Kobe University, Kobe, Japan
\IEEEauthorrefmark{2}The Asahi Shimbun Company, Tokyo, Japan
\IEEEauthorrefmark{3}Dentsu, Tokyo, Japan\\
\IEEEauthorrefmark{4}Dentsu Digital, Tokyo, Japan
\IEEEauthorrefmark{5}Human Magic, Inc., Tokyo, Japan
\IEEEauthorrefmark{6}University of Hyogo, Kobe, Japan
}
}

\maketitle

\begin{abstract}
Acquiring diverse perspectives and forming informed opinions on societal issues are essential for critical thinking and informed decision-making. Although observing debates between opposing viewpoints can promote perspective acquisition, conventional debates require substantial time and human resources, limiting their accessibility. This study investigates whether brief AI-generated rap battle-style debates can support perspective acquisition and opinion formation on societal issues. Rap battles are debate-style performances in which opposing viewpoints are expressed through rhymed verses, making them engaging while preserving the structure of argumentative dialogue. To conduct this investigation, we automatically generated rap battle-style debates using GPT-5.4-mini and presented them to participants with synthesized audio through a web-based viewing interface. We evaluated the educational effectiveness of this approach through a user study involving six participants who viewed 40 debates generated from topics on societal issues used in the All Japan Junior and Senior High School Debate Championships. The results showed that watching approximately 1.5 minute rap battle-style debates nearly doubled the number of opinions and discussion points identified by participants. Furthermore, participants reconsidered or changed their stances in 52.5\% of the evaluated cases. These findings demonstrate that even brief AI-generated rap battle-style debates can effectively support perspective acquisition and opinion formation on societal issues.
\end{abstract}


\section{Introduction}
In recent years, advances in personalized recommendation algorithms for users' preferences have reduced opportunities to encounter diverse opinions and unfamiliar topics~\cite{essay78500}.
This problem is called the filter bubble~\cite{pariser2011filter}.
It has raised concerns that users are increasingly influenced by biased content.
This makes it difficult to gain fair and balanced understanding and to access diverse opinions.

To address this issue, one effective approach is parliamentary debate (impromptu debate), which involves impromptu discussions between affirmative and opposing sides on a given topic.
It has been reported that participating in such debates can enhance understanding of diverse opinions.
Even if one is listening to debates, it can enhance one's understanding of diverse opinions.
Some findings demonstrate how listening to debates fosters a broader perspective-taking and a deeper understanding of diverse viewpoints.
Budesheim et.al reported that students who observed debates (not just participants) showed a significant increase in open-mindedness and awareness of alternative views, even when the debate challenged their initial beliefs~\cite{Budesheim99}.
Chambers advocated that, even as a passive audience, public debates can foster deliberative capacities, such as understanding opposing arguments and revising one's views in light of others~\cite{chamber2003}.

\begin{figure}[tb]
\begin{center}
\includegraphics[width=1.0\linewidth]{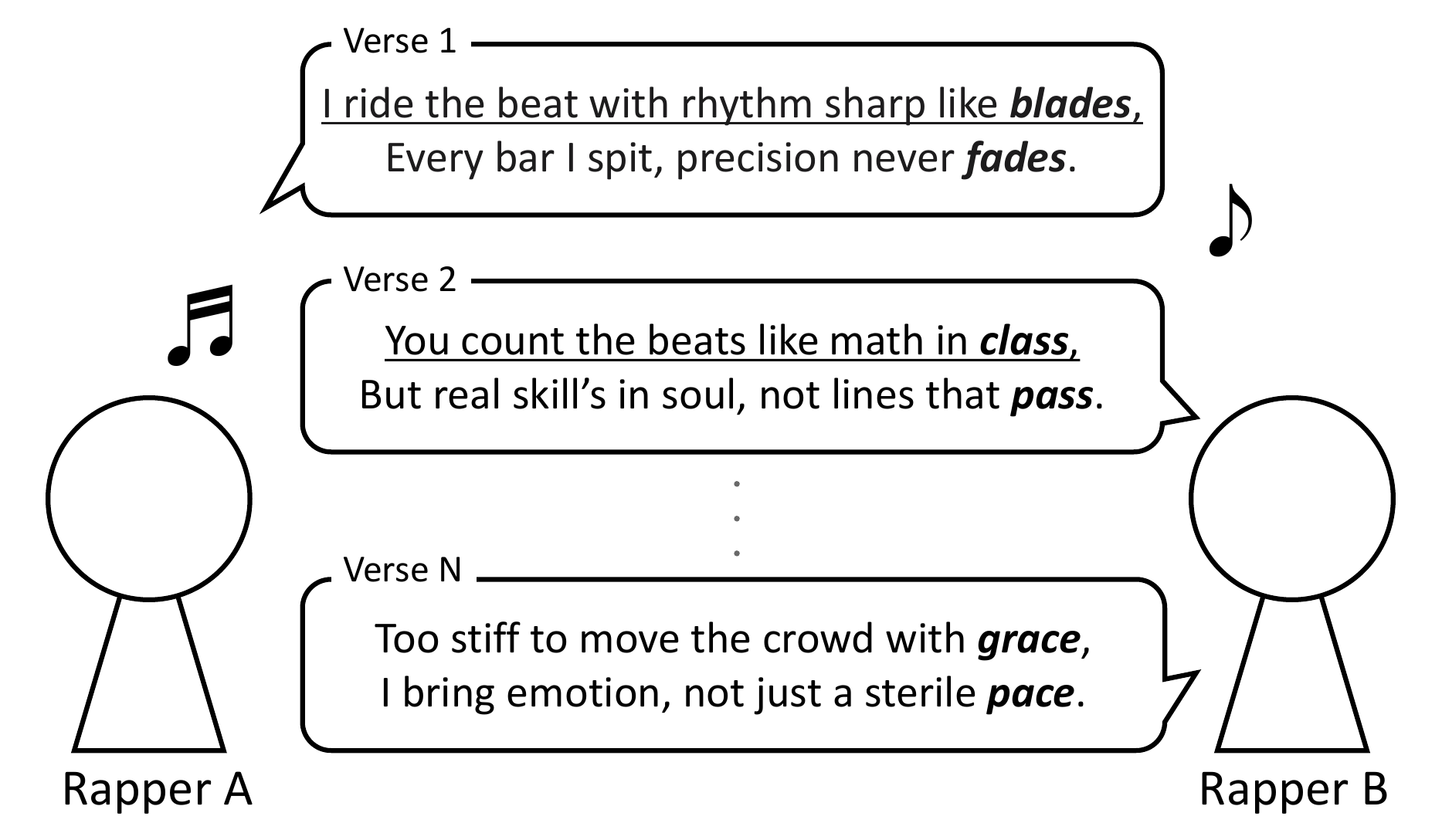}
\caption{Overview of a Rap Battle.}
\label{fig:rap_battle_img}
\end{center}
\vspace{-2.0em}
\end{figure}

However, impromptu debates typically require several tens of minutes per topic, making them relatively costly in terms of viewing time.
If short debates on various topics could be automatically generated and viewed, it would significantly reduce the time cost and enable viewers to gain a wide range of perspectives across diverse topics.

In this study, we focus on rap battles, a competitive format similar to parliamentary debates, and explore their use for generating debates.
As illustrated in Figure~\ref{fig:rap_battle_img}, a rap battle is a contest in which two rappers take turns performing improvised rap verses.
These verses typically include rhymes and responses to the opponent's lines.
Due to their short duration, concise structure, and the unique appeal of improvised lyrical exchanges, rap battles attract a large number of viewers~\cite{mibayashi2023rapbattle}.
Moreover, it has been reported that debate formats involving competition and judgment also have educational benefits, suggesting that rap battles may serve as an engaging and low-cost method to expose audiences to diverse perspectives.

To this end, we investigate whether AI-generated rap battle-style debates can facilitate the acquisition of diverse perspectives and opinion formation through debate viewing. For this purpose, we generated rap battle-style debates using GPT-5.4-mini while preserving the consistency of opposing arguments, and presented them with synthesized rap audio through a web-based viewing interface. We then evaluated their educational effectiveness through a user study.

For the evaluation, we generated 40 debates based on topics from the All Japan Junior and Senior High School Debate Championships. Six participants viewed the generated debates, with each participant assigned to one of two sets of 20 topics. 
The results demonstrate that the proposed approach effectively supports perspective acquisition and opinion formation while maintaining learner engagement and comprehension through the rap battle format.

Our main contributions are summarized as follows.
\begin{itemize}
\item
We present the first AI-based framework for automatically generating and presenting rap battle-style debates, enabling debate content to be delivered in a short and engaging format.
\item
Through a user study involving 40 debate topics, we demonstrate that brief AI-generated rap battle-style debates effectively support perspective acquisition and opinion formation by nearly doubling the number of identified opinions and discussion points and prompting participants to reconsider or change their stances in 52.5\% of the evaluated cases.
\item
Our results suggest that the rap battle format maintains learner engagement and supports comprehension, highlighting its potential as an effective medium for presenting educational debate content.
\end{itemize}

\section{Related Work}
Previous approaches to rap lyric generation have explored rhyme-aware techniques, such as replacing words in lyrics to improve rhyming~\cite{mikolov2013distributed}, and learning rhyme patterns directly from lyric corpora~\cite{potash2015ghostwriter, malmi2016dopelearning, manjavacas2019generation, xue2021deeprapper}.
Several studies have also focused on rap battle-specific generation, employing models based on LSTMs or Transformers~\cite{vaswani2017transformer}, such as in~\cite{mibayashi2023rapbattle, wu2015rapbattle, Savery2020shimon}.
More recently, methods that utilize large language models (LLMs) to generate diverse opinions on given topics have also been proposed~\cite{Hayati2024Diverse}.
In this study, we propose a generation method based on GPT-5.4-mini using in-context learning~\cite{brown2020language}, aimed at producing rap battles that incorporate both rhyme and coherent debate content.

\section{Debate Viewing System}

\subsection{Rap Battle Format Debate Generation}
We used OpenAI's API for debate generation, employing the GPT-5.4-mini model\footnote{\url{https://developers.openai.com/api/docs/models/gpt-5.4-mini}}.
Prompts were created in Japanese using in-context learning~\cite{brown2020language}, providing GPT-5.4-mini with constraints for conducting debates in a rap battle format and presenting one exchange as a one-shot example.

Following the general structure of a typical rap battle, each generated debate consisted of two exchanges, resulting in a total of four improvised rap verses.
Each verse consisted of eight lines.
To support accessibility, the model was also instructed to generate English translations of the debate and assign MC names to the two rappers.
The prompt used for debate generation is shown below.
GPT-5.4-mini was initialized with the prompt, after which debates were generated based on topics.

\begin{promptbox}
\textbf{Prompt:}
\footnotesize
Please generate a debate based on the given topic.The debate should take the form of a rap battle between two rappers.Each rapper should represent either the affirmative or the opposing side.Since the rap battle is between two personas with opposing viewpoints,``Verse 1A'' and ``Verse 2A'' must consistently argue from one side,and ``Verse 1B'' and ``Verse 2B'' must consistently argue from the other side.Each second verse should continue on the same topic,but avoid repeating the same arguments from the first verse.Each verse must consist of 8 lines,and each line should be labeled clearly as ``Line 1)'', ``Line 2)'', and so on.Assign an English MC name to each rapper representing a stance,and display the name at the beginning in brackets 【 】.For example, names like 【LandLord】 or 【FlexFlow】 are possible, but do not reuse these exact names.\\

Output Format:\\
Generate each verse in Japanese, and also provide the corresponding English translation. Please keep each line to approximately 15 characters and avoid making lines excessively long. Use the following template to generate the content:\\

Template:\\
First rapper: 【Rapper A (insert MC name)】\\
Second rapper: 【Rapper B (insert MC name)】\\
Verse 1A (rap from Rapper A's perspective):\\
Line 1) ``Japanese line / English translation``\\
...\\
Line 8) ``Japanese line / English translation``\\
Verse 1B (rap from Rapper B's perspective, opposing Verse 1A):\\
Line 1) ``Japanese line / English translation``\\
...\\
Line 8) ``Japanese line / English translation``\\
Verse 2A (rap from Rapper A's perspective, countering Verse 1B):\\
Line 1) ``Japanese line / English translation``\\
...\\
Line 8) ``Japanese line / English translation``\\
Verse 2B (rap from Rapper B's perspective, countering Verse 2A):\\
Line 1) ``Japanese line / English translation``\\
...\\
Line 8) ``Japanese line / English translation``\\

Please incorporate rhymed expressions, which are essential in rap, as shown in the example below.\\
Example:\\
Verse 1A:\\
Line 1)エーアイで描く新たな視界 (AI draws a new vision in sight)\\
...\\
Line 4)古い手法は もう捨て去る(Old methods are now discarded)\\

Verse 2B:\\
Line 1)機械に頼れば 魂消える (Relying on machines makes the human spirit fade)\\
...\\
Line 4)人間の力 今こそ示す (Now is the time to show the power of humanity)
\end{promptbox}
\label{fig:prompt_rhyme}

\begin{figure*}[t]
\centering
\includegraphics[width=0.72\linewidth]{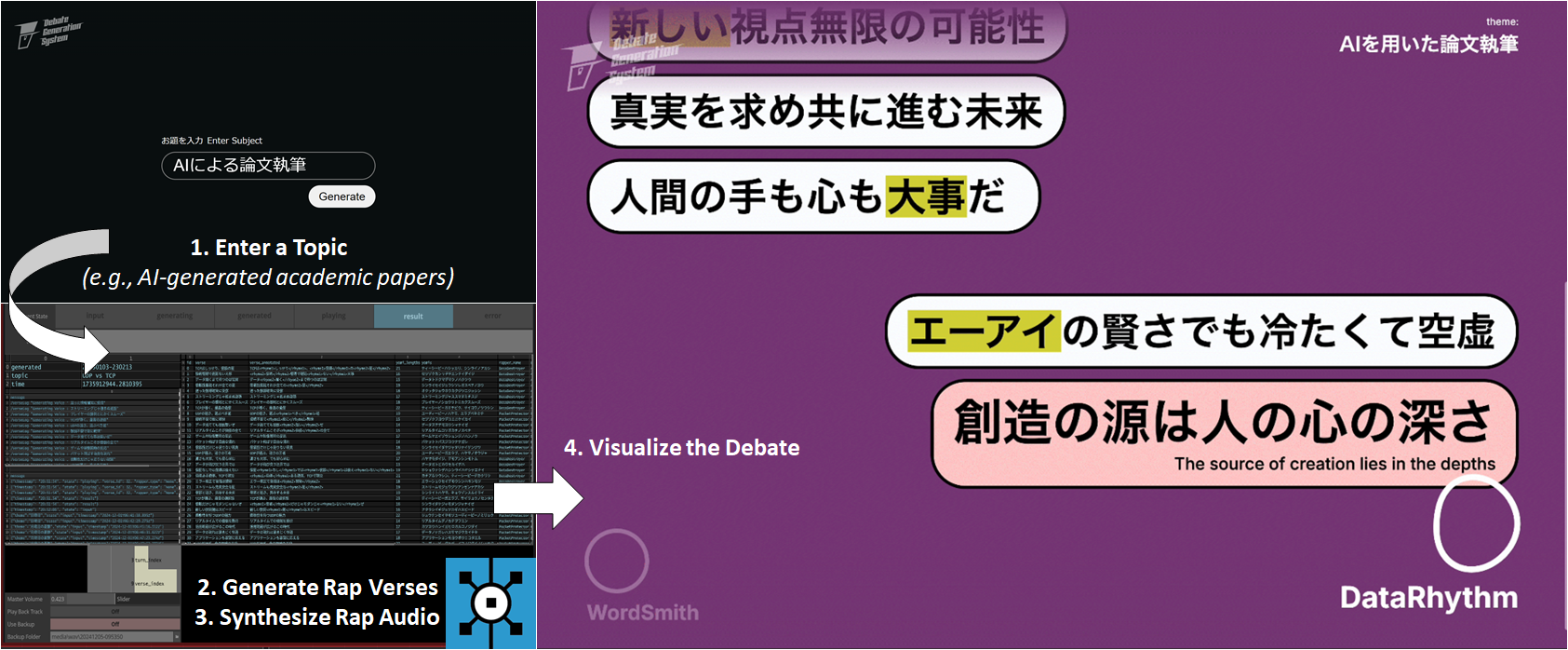}
\caption{Overview of the proposed debate viewing system. A user enters a debate topic, the system automatically generates rap battle debates and speech, and the resulting debate is visualized for playback.}
\label{fig:demo_system}
\end{figure*}

\begin{table*}[tb]
\centering
\caption{Questionnaire items and response formats used in the user study.}
\label{tab:questionnaire}
\begin{tabular}{c|p{10.5cm}|p{2.8cm}}
\hline
Item & Question & Response format \\
\hline
Q1-1 & How familiar are you with this topic? 
     & 5-point Likert scale \\
Q1-2 & What is your current stance on this topic? 
     & 6-point stance scale \\
Q1-3 & List possible opinions or discussion points regarding this topic. 
     & Free-text response \\
\hline
Q2-1 & Did you learn new opinions or discussion points through the debate? 
     & 5-point Likert scale \\
Q2-2 & What is your stance on this topic after watching the debate? 
     & 6-point stance scale \\
Q2-3 & Did the debate make you reconsider your opinion on this topic? 
     & 5-point Likert scale \\
Q2-4 & List any new opinions or discussion points you learned from the debate. 
     & Free-text response \\
\hline
Q3-1 & Was the debate natural as a rap performance? 
     & 5-point Likert scale \\
Q3-2 & Did the rap format preserve the meaning of the debate? 
     & 5-point Likert scale \\
Q3-3 & Did the debate include rhymes? 
     & 5-point Likert scale \\
Q3-4 & Did the debate include responses (answers) between the speakers? 
     & 5-point Likert scale \\
\hline
\hline
QS-1 & Was the video interface easy to watch? 
   & 5-point Likert scale \\
QS-2 & Did the rap battle format make the debate easy to understand? 
   & 5-point Likert scale \\
QS-3 & Did the rap battle format help you understand the discussion? 
   & 5-point Likert scale \\
QS-4 & Did the rap battle format keep you engaged until the end? 
   & 5-point Likert scale \\
QS-5 & Would you like to use this system to watch debates on other topics? 
   & 5-point Likert scale \\
\hline
\end{tabular}
\end{table*}

\subsection{Rap Battle Format Debate Viewing System}
Building on the prototype presented in~\cite{mibayashi2024debate}, we developed a web-based debate viewing system for the user study. The system allows users to input a debate topic, automatically generates a rap battle-style debate using the proposed method, synthesizes the rap audio, and presents the generated debate through an integrated web interface.

As shown in Figure~\ref{fig:demo_system}, the user first enters a debate topic through the web interface. The system then embeds the topic into the previously described initial prompt and generates a debate using GPT-5.4-mini.
The generated debate is then converted into rap audio via speech synthesis using Style-Bert-VITS2.
Style-Bert-VITS2 was fine-tuned using rap audio data provided by two rappers. The text encoder component employed a pre-trained DeBERTa~\cite{he2021deberta} model released by Kyoto University~\footnote{\url{https://huggingface.co/ku-nlp/deberta-v2-large-japanese-char-wwm}}.
Finally, the generated rap audio and corresponding text were visualized in a web browser in a format replicating the flow of an actual rap battle, with rhyming phrases highlighted in color.

The entire process from topic input to debate generation and visualization can be completed in approximately 20 seconds.
Since each debate video is approximately 1.5 minutes long, the entire process, from debate generation to viewing, can be completed in approximately two minutes.
This demonstrates that debates can be generated and viewed at significantly lower cost in terms of human effort and time, compared to actual parliamentary debates.

\begin{figure*}[tb]
\begin{center}
\includegraphics[width=0.75\linewidth]{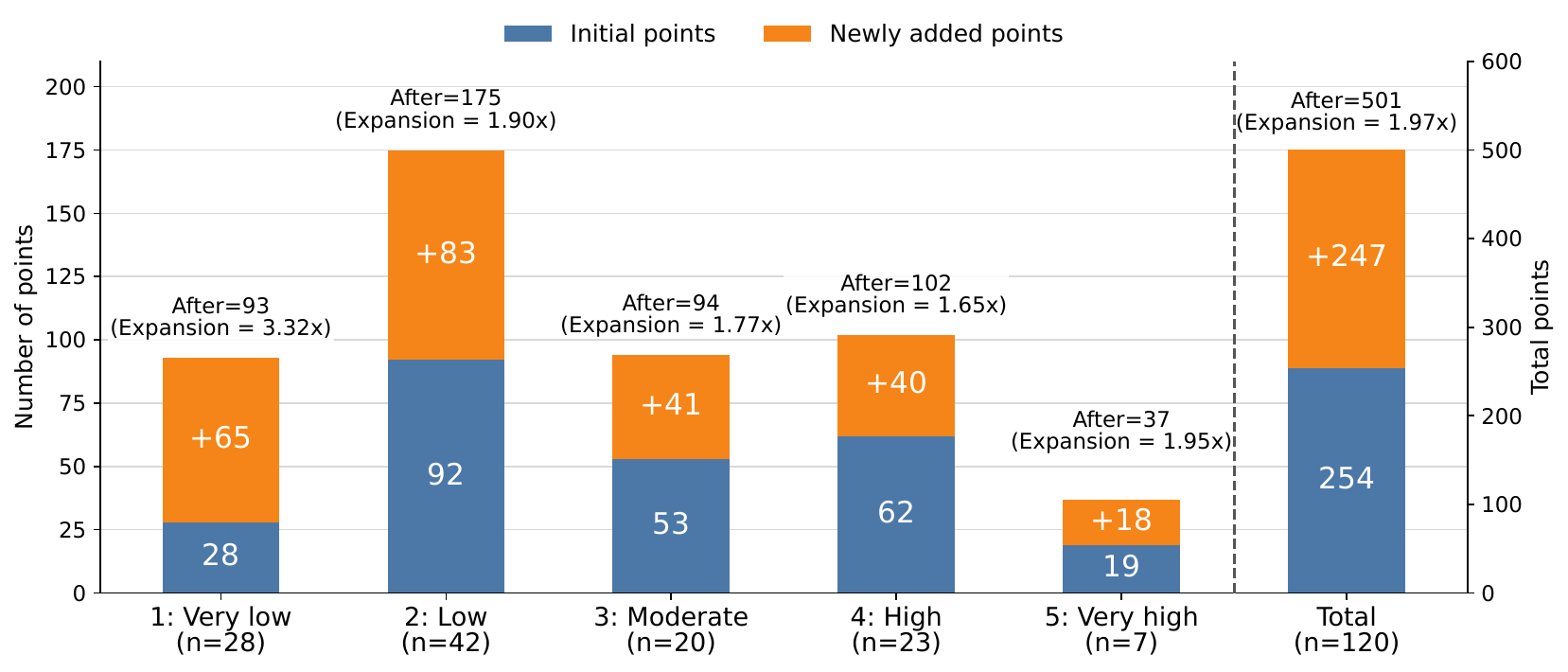}
\caption{Number of newly reported opinions and discussion points by prior topic understanding.}
\label{fig:growth}
\end{center}
\vspace{-1.5em}
\end{figure*}

\section{Evaluation}

\subsection{Experimental Setup for the Evaluation}
For the evaluation, we collected all past debate topics on societal issues used in the \textit{All Japan Junior and Senior High School Debate Championships}\footnote{\url{https://nade.jp/koshien/rondai-history/}}, resulting in 40 unique topics.
For each topic, the participants answered evaluation questions before and after watching a generated rap battle-based debate using the proposed viewing system.
An anonymized playlist of the evaluation videos is provided as supplementary material\footnote{\url{https://www.youtube.com/playlist?list=PLNf-EcxpTF9M}}.

The evaluation consisted of 5-point Likert-scale ratings, a 6-point stance scale, and free-text responses.
As shown in Table~\ref{tab:questionnaire}, the evaluation included 16 items grouped into four categories: prior understanding (Q1), educational effects of the debates (Q2), quality of the rap battle-based debates (Q3), and overall system evaluation (QS).
For Q1-2 and Q2-2, participants selected one of six stance categories before and after watching each debate: Agree (A), Somewhat Agree (SA), Neutral (N), Somewhat Disagree (SD), Disagree (D), or Undecided (U).
For Q1-3 and Q2-4, participants entered each opinion or discussion point on a separate line.
In contrast, the QS items were answered only once after participants had completed the evaluation of all topics.

The participants consisted of four graduate students and two undergraduate students, all of whom were native speakers of Japanese.
The participants were divided into two groups, and each group was assigned a different set of 20 topics.
Each set was constructed to include an equal number of junior high school and high school debate topics (10 each).
Each group consisted of two graduate students and one undergraduate student.
Each evaluator independently evaluated the 20 topics assigned to their group.

\subsection{Acquisition of New Opinions and Discussion Points}
Using the responses to Q1-1, Q1-3, and Q2-4, we compared the numbers of opinions and discussion points identified by participants before and after watching the debates.
As shown in Figure~\ref{fig:growth}, participants identified 254 opinions and discussion points before watching the debates and reported an additional 247 new opinions and discussion points afterward.
This result indicates that participants acquired nearly as many new opinions and discussion points as they had originally identified, nearly doubling the total number of opinions and discussion points.
This finding is also consistent with the subjective evaluation, as Q2-1 received an average score of 3.82, indicating that participants perceived that they had acquired new opinions and discussion points through the debates.

Furthermore, when the results were analyzed according to participants' prior understanding, the ``Very low'' group exhibited the largest increase, with the number of opinions and discussion points reaching 3.32 times the original amount.
On the other hand, participants in the ``Very high'' group also reported additional opinions and discussion points, indicating that even participants with sufficient prior knowledge were able to acquire new perspectives through the debates.
These findings suggest that the proposed system can promote awareness of diverse perspectives regardless of participants' prior understanding and has the potential to support multidimensional thinking and opinion formation.

\begin{figure}[tb]
\begin{center}
\includegraphics[width=0.7\linewidth]{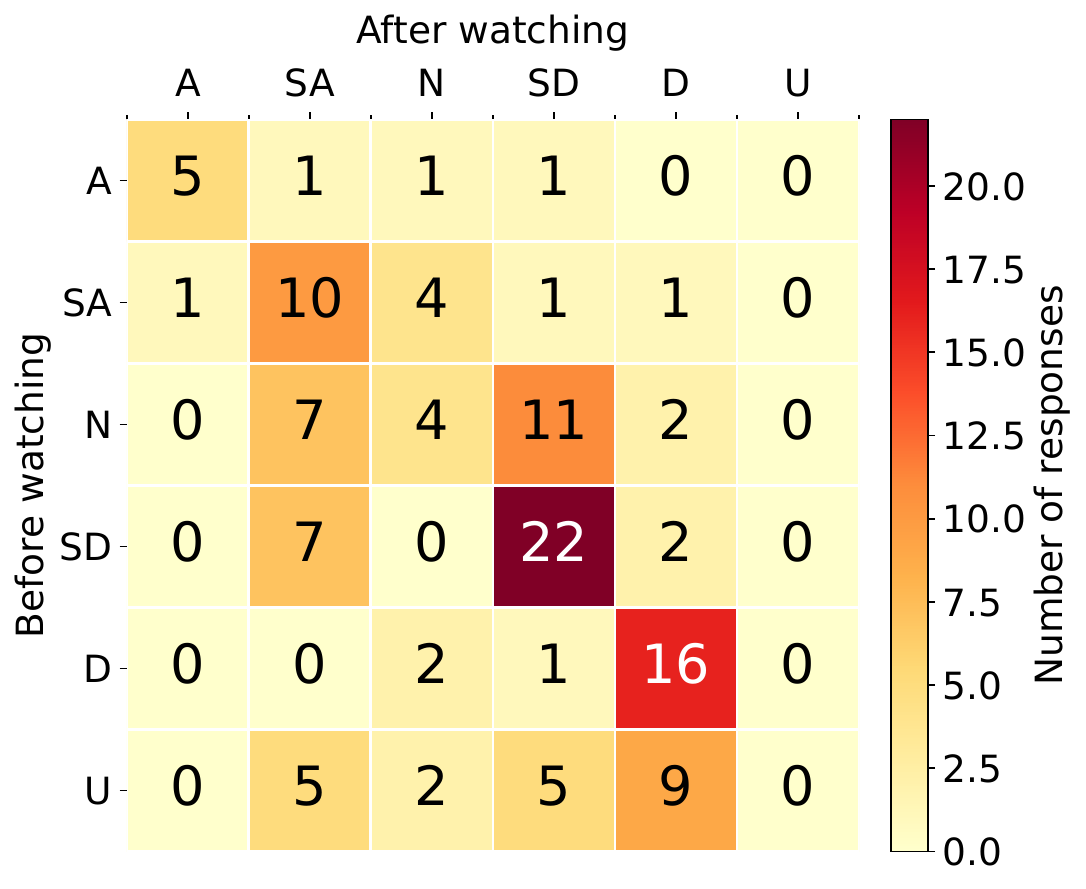}
\caption{Transitions in participants' stances between Q1-2 (before watching) and Q2-2 (after watching). A: Agree, SA: Somewhat Agree, N: Neutral, SD: Somewhat Disagree, D: Disagree, U: Undecided.}
\label{fig:stance_changes}
\end{center}
\vspace{-2.0em}
\end{figure}

\begin{table*}[tb]
\centering
\caption{Generated rap battle for the topic “Should Japan introduce fees for ambulance use?” ($n$ denotes the turn number).}
\label{tab:rap_battle}
\begin{tabular}{|p{0.3cm}|p{8cm}|p{8cm}|}
\hline
$n$ & \textbf{\textit{MC Economic} (First Rapper)} & \textbf{\textit{MC Freedom} (Second Rapper)} \\ \hline
1 & \begin{enumerate}
\item 有料化すべき救急車のシステム \newline (We should charge for ambulance system)
\item 無駄な利用を止めるためのウィズダム \newline (Wisdom to stop unnecessary usage)
\item コスト意識で使い方が変わる \newline (Cost consciousness will change the usage)
\item 平等に貢献無駄は消えてさる \newline  (Contribute equally, waste will vanish and part)
\item 予算節約病院も助かる \newline  (Saving budgets, hospitals get help)
\item 限りある資源正しく使うため \newline  (Use limited resources correctly, under same temple)
\item 社会の未来これで守れる \newline (Protect society's future, this way we sample)
\item 有料化で健全化これが我らの夢 \newline (Make it healthy with charge, this is our dream ample)
\end{enumerate} &
\begin{enumerate}
\item 有料化なんて誰が得する？ \newline (Who gains from charge, do you even bother?)
\item 命の救助は金では図れない \newline (Lives can't be measured, by nickels or dollars)
\item 貧しき者が使えなくなる \newline (The needy will suffer, a thought that appalls)
\item 緊急時に誰もが平等に \newline (In emergencies, equality's the true call)
\item 制度のハードル高くしないで \newline (Don't raise the barriers, don't make them tall)
\item 人としてのやさしさを忘れず \newline (Never forget kindness, as humans, we stand tall)
\item 負担を減らす他の方法を探せ \newline (Seek another way to lessen the load we haul)
\item 財布じゃなくて命守るのが希望 \newline (Hope is to protect life, not the purse, above all)
\end{enumerate} \\ \hline
2 &
\begin{enumerate}
\item 無駄な呼び出し減らせるこの策 \newline (Reduce these false calls, this is our track)
\item 軽度の症状で呼んじゃダメさ \newline (Don't call for mild symptom, that's the plaque)
\item 重症者のため道を開ける \newline (For serious cases, open the freeway sack)
\item 有料化で管理の透明性 \newline (Transparency of management, make no lack)
\item 割引制度で配慮も可能 \newline (Discounts available, sure we adapt)
\item 本当の緊急時は例外とする \newline (In real emergencies, exceptions unpack)
\item 正義ための合理化必要なファクター \newline (Rational for justice, the needed factor's stack)
\item 未来に繋ぐ経済の道筋 \newline (Economic future's path, this is our pact)
\end{enumerate} &
\begin{enumerate}
\item 無駄と決めるのは誰の権利だ？ \newline (Who has the right to label what's slack?)
\item 命の重みお金で測る？ \newline (Can you measure life's worth in mere stacks?)
\item 緊急の定義きちっと決めよう \newline (Let's define what's urgent, no going back)
\item 金額で邪魔するのは不服だ \newline (It's unsatisfactory, this charge to attack)
\item 人々の不安これで増すだけ \newline (This only raises worry, it's like a knack)
\item 健康のため移行すべきは違う道 \newline (For health's sake, shift on different track)
\item 本当に救急を要する時に \newline (In times truly needing ambulances stack)
\item 制度の本質を見直そうや \newline (Let's revamp the core, reflect and retract)
\end{enumerate} \\ \hline
\end{tabular}
\end{table*}

\begin{table}[tb]
\centering
\caption{Average evaluation scores for rap-related aspects of the generated debates. All questions were rated on a 5-point scale (1: strongly disagree, 5: strongly agree).}
\label{tab:rap_evaluation}
\begin{tabular}{l|l|c|c|c}
\hline
Item & Question & Group A & Group B & Overall \\ \hline
Q3-1 & Naturalness as rap & 4.28 & 4.52 & 4.40 \\
Q3-2 &  Preservation of meaning & 4.32 & 4.77 & 4.54 \\
Q3-3 &  Inclusion of rhymes & 3.70 & 4.17 & 3.93 \\
Q3-4 &  Inclusion of responses & 4.65 & 4.67 & 4.66 \\
\hline
\end{tabular}
\end{table}

\subsection{Effect on Stance Formation}
Using the responses to Q1-2 and Q2-2, we analyzed the effect of watching the debates on participants' stances.
Among the 120 responses (20 topics × 6 participants), 63 (52.5\%) showed a change in participants' stances after watching the debates.
This result is also consistent with the participants' subjective evaluations, as Q2-3 received an average score of 3.09, suggesting that the debates encouraged participants to reconsider their opinions to some extent.

On the other hand, complete reversals of stance were relatively rare.
Rather than shifting directly from strong agreement to opposition or from strong opposition to agreement, participants more often moved to adjacent categories or toward weaker or more neutral positions. 
For example, transitions such as N→SD (11), SD→SA (7), N→SA (7), and U→D (9) were frequently observed, indicating that participants tended to refine or moderate their views rather than completely reverse them.
Notably, many participants maintained their original opposing stance, with SD→SD (22) and D→D (16) being the most common transitions among those who initially disagreed.
These results suggest that the proposed debate system encourages thoughtful reconsideration of participants' stances without inducing excessive or abrupt opinion shifts.

In particular, ``Undecided'' was selected in 21 cases before watching the debates, and all of these cases shifted to another stance afterward.
Among these, 19 cases (90.5\%) shifted to either an agreeing or a disagreeing stance, while the remaining 2 cases (9.5\%) shifted to a neutral stance.
These findings indicate that the proposed system effectively supported participants in forming their own stance on the discussion topics.

\subsection{Evaluation of Rap Quality}
We evaluated the generated debates from the perspective of rap-related expression.
As shown in Table~\ref{tab:rap_evaluation}, all evaluation items received relatively high average scores.
\textit{Inclusion of responses} received the highest score (4.66), indicating that the generated debates successfully maintained interactive exchanges between opposing speakers.
\textit{Preservation of meaning} (4.54) and \textit{Naturalness as rap} (4.40) also received high scores, suggesting that the generated debates preserved the original discussion content while remaining natural as rap performances.
In contrast, \textit{Inclusion of rhymes} received the lowest score (3.93), suggesting that rhyme generation remains the primary area for improvement.

\begin{table*}[tb]
\centering
\caption{Average evaluation scores for the debate viewing system. All questions were rated on a 5-point Likert scale (1: strongly disagree, 5: strongly agree).}
\label{tab:system_evaluation}
\begin{tabular}{c|p{9cm}|c|c|c}
\hline
No. & Question & Group A & Group B & Overall \\ 
\hline
QS-1 & Was the video interface easy to watch?  & 4.67 & 4.67 & 4.67 \\
QS-2 & Did the rap battle format make the debate easy to understand?  & 4.33 & 3.67 & 4.00 \\
QS-3 & Did the rap battle format help you understand the discussion?  & 4.67 & 4.00 & 4.33 \\
QS-4 & Did the rap battle format keep you engaged until the end?  & 4.67 & 4.67 & 4.67 \\
QS-5 & Would you like to use this system to watch debates on other topics?  & 5.00 & 4.33 & 4.67 \\
\hline
\end{tabular}
\end{table*}

\subsection{Analysis of Generated Debates}

Table~\ref{tab:rap_battle} presents an example of a generated debate for the topic \textit{``Should Japan introduce fees for ambulance use?''}
The generated debate maintained clear affirmative and opposing positions throughout the interaction. For example, \textit{MC Economic} argued, \textit{``Don't call for mild symptoms, that's misuse,''} whereas \textit{MC Freedom} stated, \textit{``Can you measure life's worth in mere stacks?''} These lines reflect consistent affirmative and opposing positions, respectively, in line with the human evaluation results for stance consistency.
The debate also covered multiple perspectives. The affirmative side emphasized appropriate ambulance use and prioritizing emergency patients, as illustrated by \textit{``Make way for those in critical need.''} In contrast, the opposing side focused on equal access to emergency services, stating, \textit{``The needy will suffer, a thought that appalls.''} Such contrasting viewpoints may help viewers acquire diverse perspectives on the topic.
Furthermore, the second turn demonstrated relevant responses to the opponent's arguments. For example, the claim \textit{``Reduce these false calls, this is our track''} was followed by the response \textit{``Who has the right to label what's slack?''} This indicates that the generated debate went beyond repeating each rapper's own position and followed the intended rap battle structure. 
Finally, while the English translations contained end-of-line rhymes, rhyme generation in Japanese remained limited, consistent with the human evaluation results.

\subsection{Evaluation of the Debate Viewing System}
Table~\ref{tab:system_evaluation} shows the results of the human evaluation of the debate viewing system.
The overall results indicate that all evaluation questions received average scores of at least 4.00.
The high score for QS-1 (4.67) suggests that the user interface did not negatively affect the viewing experience, allowing the remaining evaluations to primarily reflect the quality of the debate content.

QS-2 (4.00) and QS-3 (4.33) indicate that the rap battle format did not hinder comprehension and instead supported participants' understanding of the discussion.
The high score for QS-4 (4.67) indicates that the rap battle format effectively maintained participants' engagement throughout the viewing experience.
Finally, the high score for QS-5 (4.67) suggests that participants were willing to continue using the proposed system for other debate topics, indicating its potential to support sustained learning across diverse discussion topics.

Overall, viewing approximately 1.5 minute rap battle-based debates nearly doubled the number of identified opinions and discussion points while changing participants' stances in 52.5\% of the evaluated cases.
These findings demonstrate that the proposed viewing system can effectively promote perspective acquisition and support stance formation within a short viewing time.

\section{Conclusions and Future Work}
In this study, we demonstrated that even brief AI-generated rap battle-style debates can effectively support perspective acquisition and opinion formation on societal issues.
A user study involving six participants showed that debates generated by GPT-5.4-mini effectively promoted perspective acquisition and opinion formation while maintaining learner engagement and supporting comprehension.
These findings suggest that AI-generated rap battle-style debates have the potential to serve as engaging educational materials for exploring multiple viewpoints on societal issues. 
Future work includes conducting larger-scale user studies to further investigate the educational impact of the proposed approach and improving rhyme quality while preserving debate consistency.

\bibliographystyle{IEEEtran}
\bibliography{references}

\end{document}